# Silicon Oxide is a Non-Innocent Surface for Molecular Electronics and Nanoelectronics Studies

Jun Yao,[1] Lin Zhong,*[2,3] Douglas Natelson,*[3,4] and James M. Tour*[2,5]

[1]Applied Physics Program through the Department of Bioengineering; [2]Department of Computer Science; [3]Department of Electrical and Computer Engineering; [4]Department of Physics and Astronomy; [5]Department of Chemistry, Rice University, 6100 Main St., Houston, Texas 77005

*Email addresses: lzhong@rice.edu; natelson@rice.edu; tour@rice.edu

**Abstract** Silicon oxide ($SiO_x$) has been widely used in many electronic systems as a supportive and insulating medium. Here we demonstrate various electrical phenomena such as negative differential resistance, resistive switching and current hysteresis intrinsic to a thin layer of $SiO_x$. These behaviors can largely mimic numerous electrical phenomena observed in molecules and other nanomaterials, suggesting that substantial caution should be paid when studying conduction in electronic systems with $SiO_x$ as a component. The actual switching can be the result of $SiO_x$ and not the presumed molecular or nanomaterial component. These electrical properties and the underlying mechanisms are discussed in detail.

**Introduction**

Because of its good insulating properties and mature technologies in fabrication, $SiO_x$ has long been used as a passive and insulating material in electronics. In the construction of typical two-terminal electronic devices, it is frequently used as a supporting substrate for a pair of planar electrodes, or as an insulating spacer between a pair of vertical ones.

Topologically, this E-SiO$_x$-E (E denotes electrode) system defines a gap structure across which material of interest can be bridged and electrically measured. Since SiO$_x$ ordinarily contributes negligibly to conduction, the measured electrical properties are solely attributed to the material of interest. Through this approach, electrical transport properties in various molecules and nanomaterials have been investigated.[1-13]

Recently, however, it has been shown that this traditionally passive SiO$_x$ can be readily converted into an electrically active material for resistive switching memories.[14] The conduction and switching occurs through voltage-driven formation and modification of a pathway of silicon (Si) nanocrystals (NCs) embedded in the SiO$_x$ matrix, with SiO$_x$ itself also serving as the source for the formation of Si NCs.[14] This mechanistic picture reveals the intrinsic property of conduction in SiO$_x$ and therefore results in electrode-independent switching in SiO$_x$.[15] While efforts have been directed toward SiO$_x$ device fabrication, performance and switching mechanisms,[14,16] little attention has yet been paid to the implications of this conduction upon other electronic systems that use SiO$_x$ as a device component. In particular, in molecular systems, the spacing between the electrodes tends to be close in order for the molecules, either monolayers or multilayers, to be bridged between the pair of electrodes. Consequently, at modest voltages, high local electrical fields are attained in the gap defined by the pair of electrodes. It is then of critical importance to determine whether the measured electrical phenomena are truly the result of molecular conduction or resulting from SiO$_x$ conduction induced by the high electrical field.

In this article, we demonstrate that negative differential resistance-like (NDR-like) current-voltage (*IV*) curves, resistive switching and current hysteresis are intrinsic to the thin layer of $SiO_x$. These behaviors can largely mimic electrical phenomena observed in molecular and nanoelectronic systems, suggesting substantial caution must be paid when studying conduction in nanoscale systems that use $SiO_x$ as an isolating layer. Starting with a plausible molecular NDR effect, we then show that this NDR-like behavior is actually from the $SiO_x$ itself. This is done by reproducing the same effect in a bare system without the molecules. Furthermore, by employing a carbon-nanotube network to mimic a nanoelectronics study, we show that the initial conduction assists and eventually evolves into $SiO_x$ conduction. The electrical behavior and properties of $SiO_x$ conduction are further discussed, providing potential guides to allow $SiO_x$ conduction to be distinguished from other nanoelectronic behaviors. It should be noted that previous studies[14,16] have mainly focused on the memory properties of $SiO_x$ with layer thicknesses >30 nm, whereas here electrical phenomena are discussed in $SiO_x$ at a thickness between 2 nm (surface native oxide) and 10 nm, closer to molecular dimensions.

**Results and Discussion**

**1. Bistable NDR-like behaviors and resistive switching**

To sandwich molecules between a pair of electrodes, vertical polySi-$SiO_x$-polySi (x ~ 2) stacking structures with diameters of 100 μm were fabricated as illustrated in Fig. 1a. Highly doped polySi layers (ρ < 0.005 Ω·cm, thickness = 70 nm) are used here as both top and bottom electrodes to exclude any effects due to motion of metal from the electrodes.[17,18] The two electrodes were spaced by a $SiO_x$ layer with a thickness of 10 nm. By etching away some portion of the $SiO_x$ at the vertical edge in a 10:1 buffered oxide etch (J. T. Baker), a vertical nanogap system was defined and molecules can be assembled in the nanogap.[7] This structure is similar to other vertical molecular systems in which $SiO_x$ layers were used as insulating and supportive spacers between the top and bottom electrodes.[1-3] 3-Aminopropyltriethoxysilane (APTES) molecular layers at an estimated thickness of 10 nm were then assembled onto the vertical $SiO_x$ surface in the nanogap, following a process described in the literature.[19] All of the electrical characterizations were performed under vacuum ($10^{-5}$ Torr) at room temperature. The successful assembling of the molecules is indicated by the increased conduction compared to that of a pair of bare electrodes (inset in Fig. 1b). During a subsequent series of consecutive voltage sweeps from -12 V to +12 V, the current level gradually increased until it reached a certain value after several voltage sweeps; during this process, NDR-like *IV* curves were persistently observed (Fig. 1b). The NDR-like behavior was approximately symmetric in the negative and positive voltage regions, and was very similar to that observed in OPE molecules assembled in metal nanogaps on a $SiO_x$ substrate.[4]

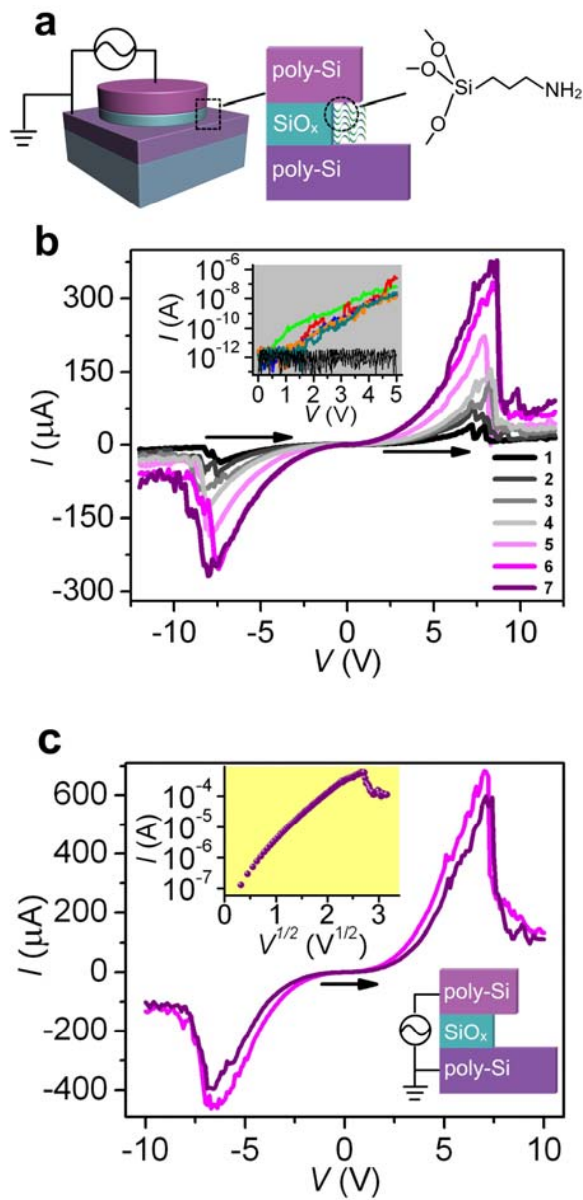

**Figure 1**. (a) Left panel: Schematic of a vertical polySi-SiO$_x$-polySi structure (70 nm-10 nm-70 nm) and the electrical-characterization setup. The diameter of the structure is 100 μm. Right panel: an enlarged schematic of the nanogap defined at the vertical edge in which APTES molecular layers are assembled. (b) Consecutive voltage sweeps (-12 V → 12 V) in a device with APTES molecular layers as shown in (a). The numbers indicate the voltage-sweep orders. Inset shows the conduction before (black curves) and after

(color curves) APTES assembling in several devices, with each curve corresponding to one device. (c) *IV* curves (-10 V → +10 V) in a bare polySi-SiO$_x$-polySi device without molecules. Note that the device was thermally annealed at 600 ºC for 10 mins in a reducing environment (Ar/H$_2$ =450/150 sccm) prior to the electrical characterization. The top inset shows one of the *IV* curves in the positive bias region in a *log(I)-V$^{1/2}$* format. The bottom inset shows a schematic of the vertical edge on which no molecules are assembled.

A control experiment in which no APTES molecules were assembled in the polySi-SiO$_x$-polySi nanogap shows almost the same NDR-like behavior (Fig. 1c). The absence of molecules here indicates that the NDR-like behavior in the previous system was not intrinsic to the molecules. Instead, it resulted from the soft breakdown of the SiO$_x$ layer at the vertical edge[14] region by the high electrical field built in the nanogap. The gradually increased current level upon continuous voltage sweeps (Fig. 1b), deviating from the initial conduction (inset in Fig. 1b), was a signature of the electroforming[20] process in SiO$_x$. In fact, similar NDR-like behaviors were first described in the 1960s in silicon-rich M/SiO$_1$/M (M denotes metal) sandwiched systems.[21] While different models were proposed for the mechanism,[22] the exclusive use of metal electrodes often led to a mechanistic picture of conduction by metal defects injected from the electrodes.[21] The defective silicon-rich SiO$_1$ system and the extrinsic metal-filamentary picture have likely contributed to the neglect of the potential conduction from SiO$_x$ in other electronic systems, where *x* is usually close to 2 to ensure a good insulating property. Indeed, a recent study[14] revealed details of SiO$_x$-generated switching: the voltage-driven electrochemical processes can induce local reduction of SiO$_x$ to form Si nanocrystalline pathways and lead to resistive switching and conduction intrinsic to $SiO_x$. Consequently, the switching is electrode-independent and can be realized in $SiO_x$ with an *x* value close to 2.

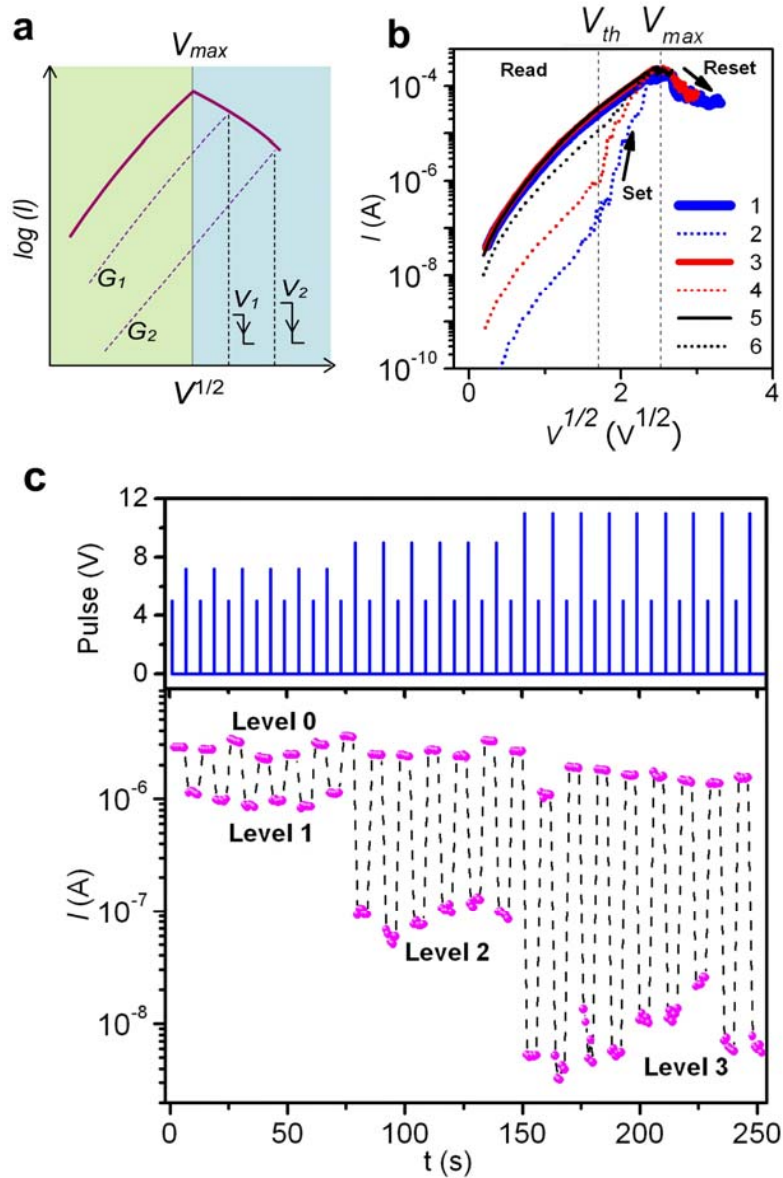

**Figure 2**. (a) An illustration of the "extrapolation rule" in a $log(I)$-$V^{1/2}$ format. (b) Consecutive *IV* curves in a bare vertical polySi-$SiO_x$-polySi device as illustrated in Fig. 1c (lower inset). The numbers indicate the voltage-sweep orders. The solid curves are voltage sweeps from 0 to a value above $V_{max}$, with blue, red, and black curves correspond to 0 V → +11 V, 0 V → +9 V, and 0 V → +7 V, respectively. The dashed curves are voltage sweeps from 0 V to a value (+6 V) close to $V_{max}$. 'Read', 'Set', and 'Reset' regions are defined by $V_{th}$ and $V_{max}$ as shown in the figure. (c) Top panel: A series of voltage-pulse sets of (+5 V, +7 V), (+5 V, +9 V), and (+5 V, +11 V) working as programming voltages. Bottom panel: corresponding memory states read by a +1 V pulses. Note the programming current is not shown here. The data indicates that a set voltage of +5 V programs the device into the ON state (level 0), whereas reset voltages of different magnitudes (+7 V, +9 V, +11 V) program it into different OFF states (Level 1, Level 2 and Level 3).

In light of the Si-pathway conduction, the NDR-like behavior can be a result of voltage-driven structural changes in the conducting pathways or filaments. For example, the conductance increase and decrease correspond to the construction and destruction of the conducting filaments, respectively. Therefore, the resistance of the $SiO_x$ depends on the history of the voltage sweeps that modified the filaments. For example, if a voltage sweeps above $V_{max}$ (the voltage at the NDR peak) and then drops fast to 0 V, the resultant resistance of the $SiO_x$ corresponds to the value at this voltage and can be estimated through an "extrapolation rule".[21] This is illustrated in Fig. 2a in which the *IV* relationship is presented in a *log(I)*-$V^{1/2}$ format, since generally the conduction in $SiO_x$ is dominated by tunneling having the characteristic of $log(I) \propto V^{1/2}$ (see upper inset in Fig. 1c).[21,22] If a voltage sweeps to $V_1$ and then drops quickly to 0 V, it is expected that the conductance of the $SiO_x$ has been changed to $G_1$ which can be estimated by drawing an extrapolation line parallel to the initial NDR curve in the $V < V_{max}$ region (Fig. 2a). In a similar way, lower conductance $G_2$ can be achieved by sweeping to a voltage $V_2$ ($V_2 > V_1$). This rule is well-demonstrated in the actual *IV* curves in Fig. 2b in which each starting conductance (dashed curves) depends on the previous voltage sweep (solid curves) through the "extrapolation rule". The *IV* curves in Fig. 2b also define the 'read', 'set', and 'reset' regions that are usually identified in resistive switching devices.[20] The 'reset" region is at $V > V_{max}$ as voltages in this region reset the $SiO_x$ to lower-conductance states. And the 'set' region begins at a threshold voltage $V_{th}$, at which the lower conductance suddenly increases and deviates from the original $log(I) \propto V^{1/2}$ curve in the 'read' region. For the above reasons, multilevel nonvolatile resistive memories can be realized by applying voltage pulses of different magnitudes (see Fig. 2c). And since the behavior implies both nonvolatile high-conductance (ON) and low-conductance (OFF) states, it can have the appearance of NDR and has been referred to as bistable NDR.[23]

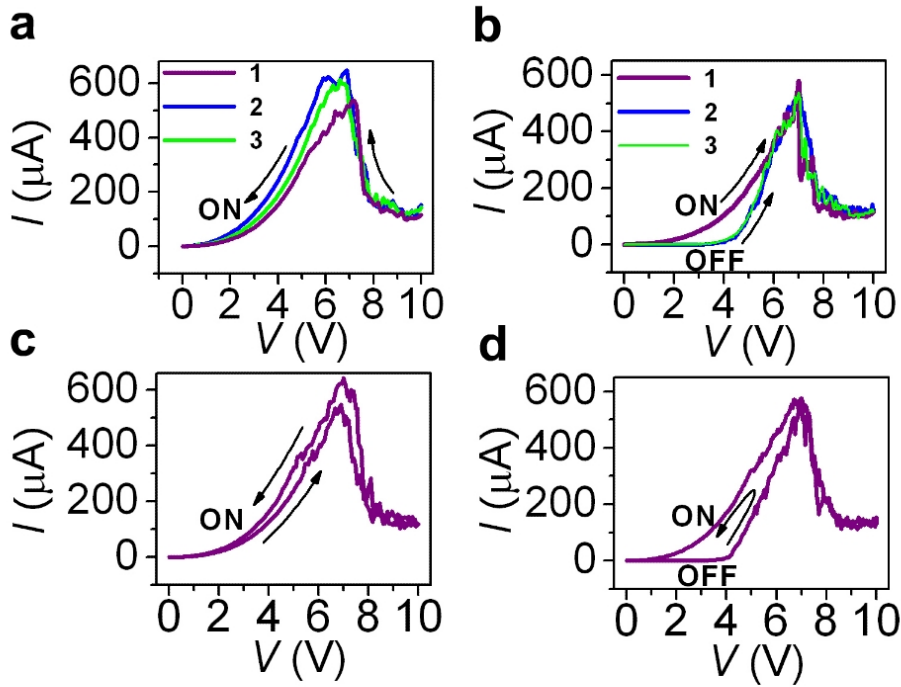

**Figure 3**. *IV* curves in a vertical polySi-SiO$_x$-polySi (70 nm-10 nm-70 nm) device by different voltage-sweep modes: (a) single backward sweeps (+10 V → 0 V); (b) single forward sweeps (0 V → +10 V); (c) Double sweep (0 V → +10 V → 0 V) starting with an ON state; (d) double sweep (0 V → +10 V → 0 V) starting with an OFF state. The arrows indicate the sweep directions and the numbers indicate the sweep orders.

For the above discussion, depending on the voltage-sweep history and mode, different *IV* behaviors can be observed in SiO$_x$. For example, for a backward voltage sweep starting from the 'reset' region to 0 V, the *IV* curve is always in an ON state in the 'read' region as it eventually bypasses the 'set' region (Fig. 3a). However, for forward voltage sweeps starting from 0 V to a reset voltage, after the first sweep, the subsequent sweeps always have OFF states in the 'read' region since each previous sweep ends at the 'reset' region (Fig. 3b). Similarly, in double-sweep modes, if initially the device is in an ON state, no current hysteresis is produced (Fig. 3c). If initially the device is in an OFF state, hysteresis is produced (Fig. 3d). This last type of sweep has been frequently used[14-16,23,24] as characteristic resistive-switching *IV* curves since it indicates both the programming regions and ON/OFF ratio. Similarly, the electroforming process that converts the pristine SiO$_x$ into a switching state can also have different *IV* evolutions with respect to different voltage-sweep modes (see Supporting Information S1).

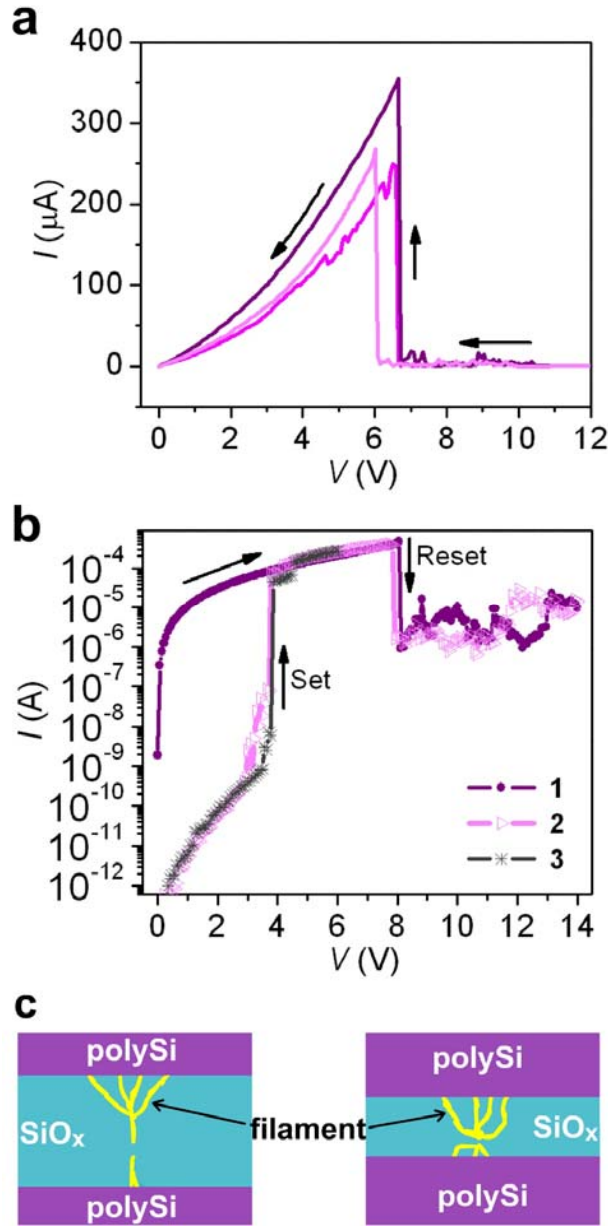

**Figure 4**. (a) Backward voltage sweeps (+12 V → 0 V) in a vertical polySi-SiO$_x$-polySi (70 nm-40 nm-70 nm) device, showing steep NDR peaks. (b) Forward *IV* curves from the same device in (a), showing sharp set and reset threshold voltages. The numbers indicate the sweep order. (c) Illustrations depicting the SiO$_x$-layer thickness effects on the development of conducting filaments and thus the OFF current. Left panel: An OFF state from a fully-developed filament in a comparatively thick SiO$_x$ layer. Right panel: An OFF state from a less effectively developed filament in a thin $SiO_x$ layer. The development of the single filamentary strand in the thick $SiO_x$ layer ensures a low OFF tunneling current, while the lack of this single filamentary strand in the thin $SiO_x$ layer results in an increased cross-section area of tunneling, thus an increased OFF current.

The bistable NDR-like peak in $SiO_x$ can be very sharp (Fig. 4a). Consequently, the resistive switching *IV* curves feature sharp set and reset threshold voltages (Fig. 4b). As the resistive switching in $SiO_x$ is through filamentary conduction, the differences in the *IV* curves were considered a consequence of different numbers of filaments.[15] For example, a sharp curve indicates few filaments, whereas a smoother curve is a collective of multitude breaking or reforming events at different voltages in various filaments. This change in the number of filaments can account for a higher ON/OFF ratio (> $10^6$ in Fig. 4b) in the sharp *IV* curves compared to that in a smooth curve (~ $10^3$, see Fig. 2b). It can also account for multilevel memory behaviors (Fig. 2c) as a higher reset voltage is expected to break more conducting filaments thus resulting in lower OFF conduction. The average ON/OFF ratios in the thin $SiO_x$ (10 nm) was typically within $10^3$, compared to ratios exceeding $10^4$ in thicker $SiO_x$ (40 nm), mainly because of increased current in the OFF states. In the mechanistic picture of an electrochemical redox process of $Si \leftrightarrow SiO_x$ at the switching site,[14] it can be understood that the effectiveness of this process, even in a single filament, can commensurately modulate the conductance. For example, a less effective oxidation process of $Si \rightarrow SiO_x$ at the switching site is expected to cause an increased OFF conduction. The dynamics of the switching process, including the degree to which redox switching of $Si/SiO_x$ can be cooperative at multiple sites, may well be affected by internal stress distributions associated with the glassy $SiO_x$ structure.[25] A restricted film thickness is likely to hinder the formation of a single filamentary strand (as illustrated in Fig. 4c), resulting in an increased cross-sectional area of tunneling in the OFF state, thus the increased OFF current. It also raises the question: At what thickness does $SiO_x$ still demonstrate the switching behavior? Our experiments indicate that reproducible bistable NDR-like behavior and resistive switching can be induced in $SiO_x$ with thicknesses ranging from 7 nm to 200 nm, covering a majority of nanogaps defined on $SiO_x$ for molecular and nanoelectronic systems that have been studied.

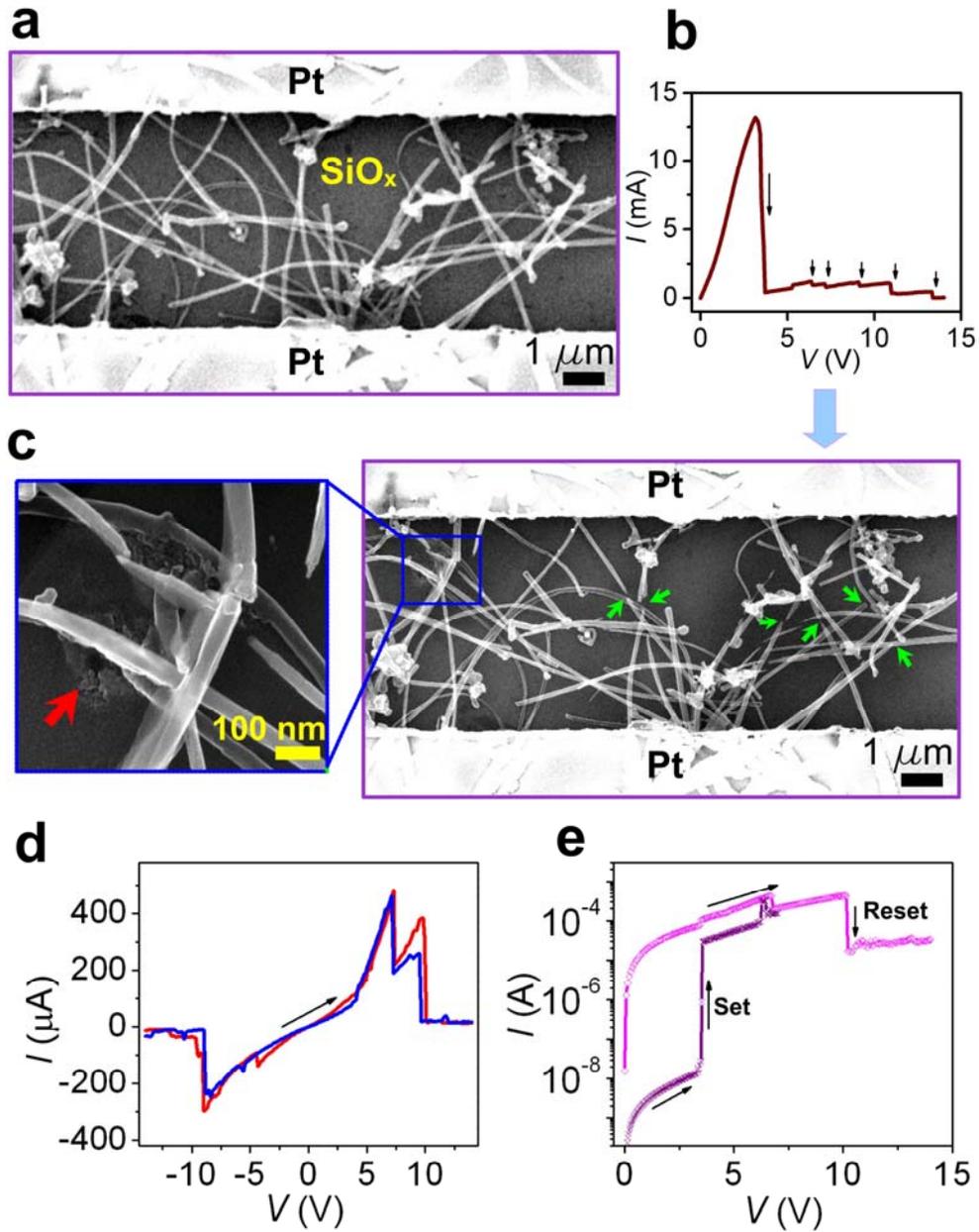

**Figure 5**. NDR-like behavior and resistive switching in a MWCNT network on a Si substrate capped with 200 nm SiO₂. (a) A scanning electron microscopy (SEM) image of the random MWCNT network. (b) Initial voltage sweep (0 V → +14 V). The black arrows indicate sudden current drops, or electrical breakdowns in MWCNTs. (c) SEM image of the same MWCNT-network device after the voltage sweep in (b). The green arrows indicate broken regions in different MWCNTs. The inset on the left panel is a zoomed-in picture of the MWCNT network. It shows the broken regions of MWCNTs, along with observable damage to the underlying $SiO_2$ substrate (indicated by the red arrow). (d). NDR-like curves in the same electroformed device. The multiple peaks are likely caused by multiple MWCNT-$SiO_x$-MWCNT switching sites. (e). Characteristic resistive switching *IV* curves in the same device.

With the assistance from other conducting pathways, this $SiO_x$ thickness could extend far beyond 200 nm. We used a network of multiwalled carbon nanotubes (MWCNTs) to mimic a random molecular layer on a $SiO_x$ substrate, and patterned the network with two electrodes over 5 μm apart (Fig. 5a). By sweeping to +14 V, multiple sudden conductance decreases are observed (Fig. 5b) as a result of electrical breakdowns in the MWCNTs. This is also visible in the physical breaking of MWCNTs (Fig. 5c, indicated by green arrows). Upon further voltage sweeps, the $SiO_x$ between certain nanogap defined by two broken ends of MWCNT can be electroformed to a resistive switching state. This process is always accompanied by visible morphological change to the $SiO_x$ at the nanogap region (left panel in Fig. 5c), which is a signature of the electroforming process in various resistive switching materials.[14,15,22] Hence, starting with a conduction initially coming from the MWCNT-network, the conduction eventually evolves into reproducible bistable NDR-like behavior and resistive switching coming from $SiO_x$ (Fig. 5d,e), with the broken MWCNT network merely serving as effective nano-spaced electrodes.

The above MWCNT network offers a vivid example of how conductive molecular layers or nanomaterials can assist the formation of $SiO_x$ switching; the disruption of molecular layers or nanomaterials, either by local electrical breakdown or nonuniformity during assembly, helps to build up a high local electrical field that leads to a soft breakdown in the $SiO_x$ layer. As the bistable *IV* behavior and resistive switching are intrinsic properties of $SiO_x$, they can be induced in the $SiO_x$ by molecules or other exogenous materials atop the $SiO_x$ substrate.[15] This $SiO_x$ soft breakdown-induced resistive switching and NDR-like behavior might be the cause of various qualitatively similar electrical behaviors in molecule layers,[4,5] carbon materials,[8-11] nanowires,[12,13] and bare nanogaps,[26-28] in which little attention, if any, was formerly paid to the $SiO_x$ substrates.

Besides building up high local fields, the initial conduction from molecules or nanomaterials also provides current local heating which could assist the electroforming in $SiO_x$ since thermal annealing was found to introduce more defects at the $SiO_x$ surface.[14,16] These introduced defects could serve as electron hopping centers so that electroforming is more easily induced at a voltage below the hard-breakdown threshold. This is also the case for the bare polySi-$SiO_x$-polySi structures (Fig. 1c) in which thermal annealing (600 ºC, 10 min, $Ar/H_2$ = 150/50 sccm) prior to electrical characterizations was adopted to facilitate the electroforming process. With a layer of APTES molecules, which introduce both enhanced local electrical field and current local heating, the system can be readily electroformed as described in Fig. 1a without thermal annealing. Not surprisingly, coating the bare polySi-$SiO_x$-polySi structure with a thin layer (5 nm) of amorphous carbon can serve the same role and leads to the electroforming of $SiO_x$ and NDR-like behavior[8,9] (see Supporting Information S2). Note that while electrical breakdown in bulk $SiO_x$ usually needs an electrical field larger than 10 MV/cm, a $SiO_x$ surface in contact with other exogenous nanomaterials, or that is subjected to thermal annealing, has more defects and is therefore an easier material in which to induce soft breakdown at a lower electrical field. Consequently, the conduction in the vertical polySi-SiO$_x$-polySi is found to be localized at the vertical SiO$_x$ edge.[14]

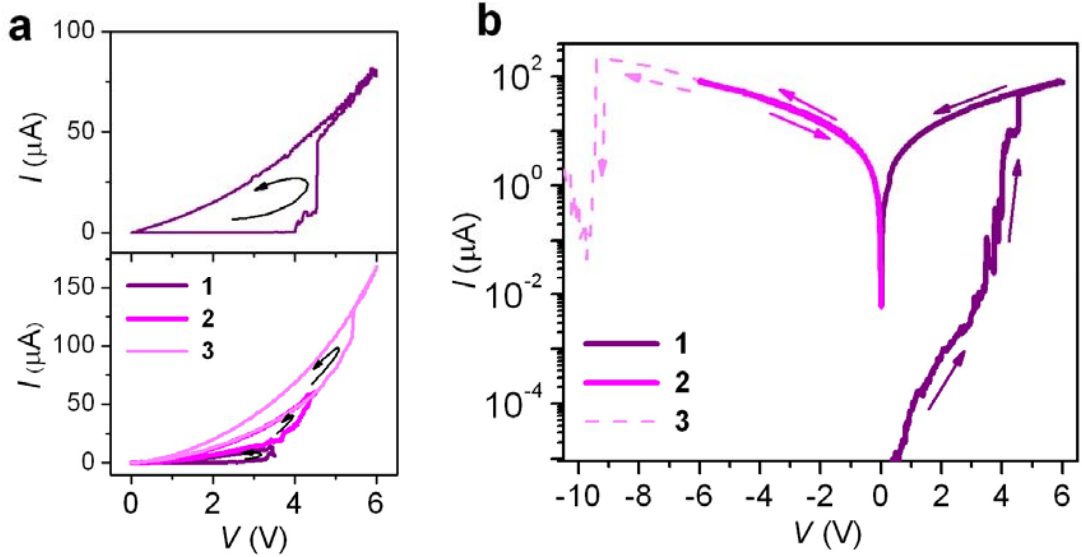

**Figure 6**. (a) Top panel: a hysteretic *IV* curve featuring a set operation in a vertical polySi-SiO$_x$-polySi device with 40 nm SiO$_x$. Bottom panel: multiple hysteretic *IV* curves featuring a series of multi-stage set operations in a vertical polySi-SiO$_x$-polySi device with 10 nm SiO$_x$. The arrows indicate the sweep directions and the numbers indicate the sweep orders. (b). *IV* sweeps in both polarities in a polySi-SiO$_x$-polySi device with 40 nm SiO$_x$. Curve 1 shows a set process (0 V → +6 V → 0 V) in the positive bias region. Curve 2 shows a subsequent voltage sweep (0 V → -6 V → 0 V) in the negative bias region, with no reset operation incurred. Curve 3 shows the next voltage sweep (0 V → -12 V) that triggers a reset operation at ~ -9 V.

While the appearance of NDR-like behavior requires a voltage sweep above $V_{max}$, the bistability infers that current hysteresis can be readily produced in the 'set' region with *V*

> $V_{th}$ (see $V_{th}$ definition in Fig. 2b). For example, starting from an OFF state, any voltage sweep to a value above $V_{th}$ incurs the set process and thus current hysteresis. The conductance increase in the hysteretic loop can be abrupt (top panel in Fig. 6a) or gradual featuring multi-stage set processes (bottom panel in Fig. 6a). These behaviors cannot only mimic current hysteresis in molecular systems,[29] but also in mechanical switches where similar vertical M-SiO$_x$-M stacking structures were adopted.[30,31] As $V_{set}$ is smaller than $V_{reset}$, this type of hysteresis is always unidirectional (counter-clockwise) toward a higher-conductance state. The non-volatility also determines that the same hysteresis is not reproducible before a reset operation is performed. In particular, a reset operation cannot be performed by sweeping to the opposite polarity of - |$V_{set}$| (Fig. 6b), but needs to sweep further to - |$V_{reset}$|. This features the typical unipolar switching that is only voltage-magnitude dependent but not polarity dependent. It should be noted that some resistive switching systems[32,33] were presented similarly in both polarities, but these are essentially unipolar behaviors as described in Fig. 4b.

The underlying cause for all the above electrical phenomena is voltage-driven formation and modification of conduction filaments (Si-NC pathways) embedded in the SiO$_x$ matrix. A certain minimum voltage is required in order to electrochemically modify the conducting pathway. Generally, the bistable NDR-like peak appears at |$V_{max}$| > 3 V, which indicates that bistable NDR-like behavior and resistive switching below 3 V are unlikely from SiO$_x$. However, there is no clear maximum $V_{max}$, since interface resistance can reduce the actual voltage drop across the SiO$_x$, thus pushing $V_{max}$ above 10 V.[15] The *IV* curve in the 'read' region ($V < V_{th}$) is comparatively smooth and dominated by tunneling. Above $V_{th}$, as voltage-driven modifications of the conducting filaments begin, current fluctuations begin. These fluctuations alone produce various local NDR-like peaks (e.g., see Fig. 6) which are not reproducible. They also persist in the electroforming process prior to the formation of bistable NDR-like behavior (Supporting Information S1). Therefore, careful attention should be paid in molecular and nanosystem characterizations, in which the reproducibility of the *IV* curves are sometimes not described or are neglected. The resistive switching in $SiO_x$ needs to be in an oxygen-deficient environment, and cannot be performed in ambient environment.[14] This may be due to the Si-filamentary nature where current local heating induced oxidation prevents the switching. However, once "programmed" by electroforming, the resistance states are air-stable with a retention time projected to be above years.[14,22] This can be a good point of difference from some charge-based molecular switching system in which the states decay faster.[7,32,34]

**2. NDR and current hysteresis at low voltage (< 3 V)**

Although the mechanisms for the bistable NDR-like behavior and resistive switching in various molecular and nanomaterial systems are largely unknown and debatable,[35] many have turned out to be through localized filamentary conduction.[35-37] In this form, the resistance change can be generally viewed as a result of electronic structural change by doping or electrochemical reactions[35] in the conducting pathways. Therefore, the conductance can be modulated in a nonvolatile manner. Meanwhile, the energy gaps between the highest molecular orbitals (HOMO) and lowest unoccupied molecular orbitals (LUMO) can produce another type of NDR behavior through resonant tunneling,[38-40] an effect similar to that in a resonant tunneling diode. Since resonant NDR results from energy-level alignment between the electrodes and the molecular orbitals modulated by an external bias, the conductance change is volatile. Also, as tunneling current decays exponentially with molecular dimensions, resonant NDR is usually observed in monolayer or few-layer molecules at low voltage bias (< 3 V) with limited currents.[41,42] These aspects differ from the behaviors in the $SiO_x$ bistable NDR-like behavior discussed above.

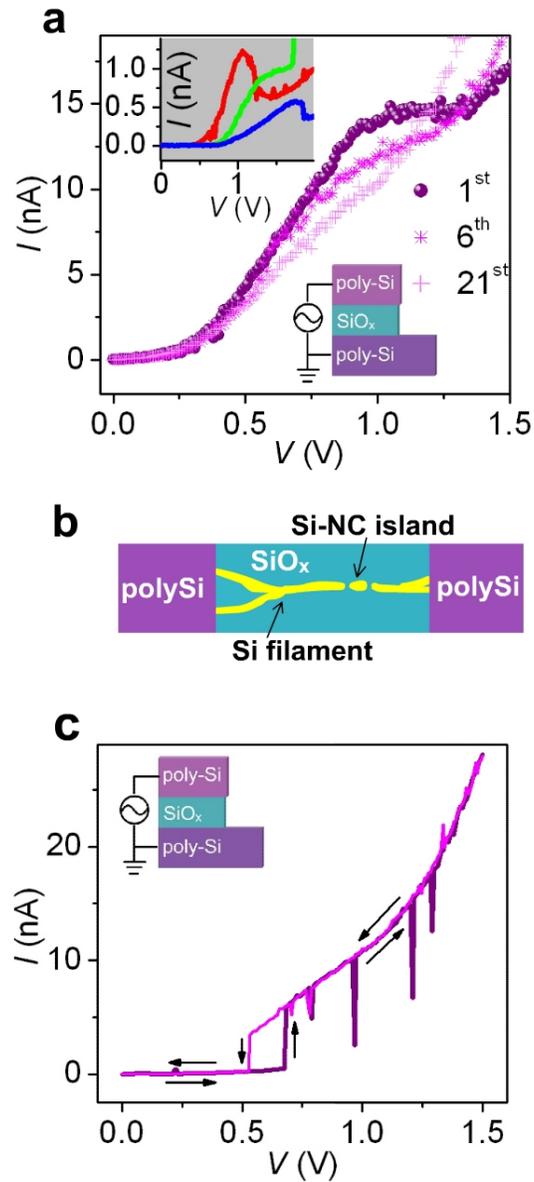

**Figure 7**. (a) NDR at low voltage in a polySi-SiO$_x$-polySi device (illustrated in the bottom inset) having the same structure and parameters as described in Fig. 1c. The numbers indicate the numbers of voltage sweeps, showing a gradual disappearance of the NDR on the 21$^{st}$ sweep. Top inset shows three different NDR curves obtained in a same device at different OFF states programmed by reset processes. (b) A schematic showing an isolated Si-NC island in the filamentary path. (c) Current hysteresis at low voltage (0

V → 1.5 V → 0 V) in a polySi-SiO$_x$-polySi device (illustrated in the top inset) having the same structure and parameters as described in Fig. 1c. The arrows indicate the voltage-sweep directions.

On the contrary, we find that at a soft-breakdown state, SiO$_x$ can also show resonant-like NDR at a low voltage region (Fig. 7a), with the NDR-peak location and current level close to those observed in molecular systems.[41,42] The appearance of this NDR can be understood based on a mechanistic picture of conduction through an aligned Si-NC pathway.[14] During the process of electroforming, or at an OFF state, the discontinuity of the Si-NCs gives rise to the possibility of forming an isolated Si-NC island along the pathway (Fig. 7b). The confinement in the Si-NC island results in discrete energy levels, thus the resonant tunneling effect. This proposed mechanistic picture is indeed supported by the experimental observation of NDR in a Si quantum dot array.[43] Since a read voltage (e.g., < 3 V) is not expected to induce structural change in the filament, this NDR is reproducible (Fig. 7a). The gradual degradation of the NDR upon continuous voltage sweeps (Fig. 7a), which was also observed in molecules,[44] may be due to charge-trap effects. As the morphology of the filament can be altered during electroforming or after different programming processes, variations in this resonant-like NDR are expected, even in the same device at different stages of the filament evolution. This is shown in the inset in Fig. 7a; three different NDR curves from the same SiO$_x$ device can be obtained after different reset processes. The ability to re-obtain the precisely desired level of resonant-like NDR behavior in SiO$_x$ is high (within 10% between NDRs in different OFF states), while the rest are through typical tunneling as described before (Fig. 2b). It should be noted that, due to the semi-analog conductance modulation through multi-stage set (Fig.

6a) or reset processes (Fig. 2c) described above, various resistance states ranging from kΩ to GΩ can be achieved in $SiO_x$. These conductions, though perhaps without NDR, can still mimic non-ohmic conduction in molecules.[2,3]

$SiO_x$ can also produce other effects in the low voltage region. Fig. 7c shows an *IV* curve with an abrupt conductance jump at ~0.65 V and, when tracking back, a sudden conductance drop at ~0.5 V, producing a hysteresis window of ~0.15 V. Similar electrical behavior has also been observed in single molecules.[45] The actual cause for this behavior in $SiO_x$ needs further investigation. It is likely that some sudden trapping and de-trapping events happen upon certain threshold voltages, thus rapidly modulating the conductance.

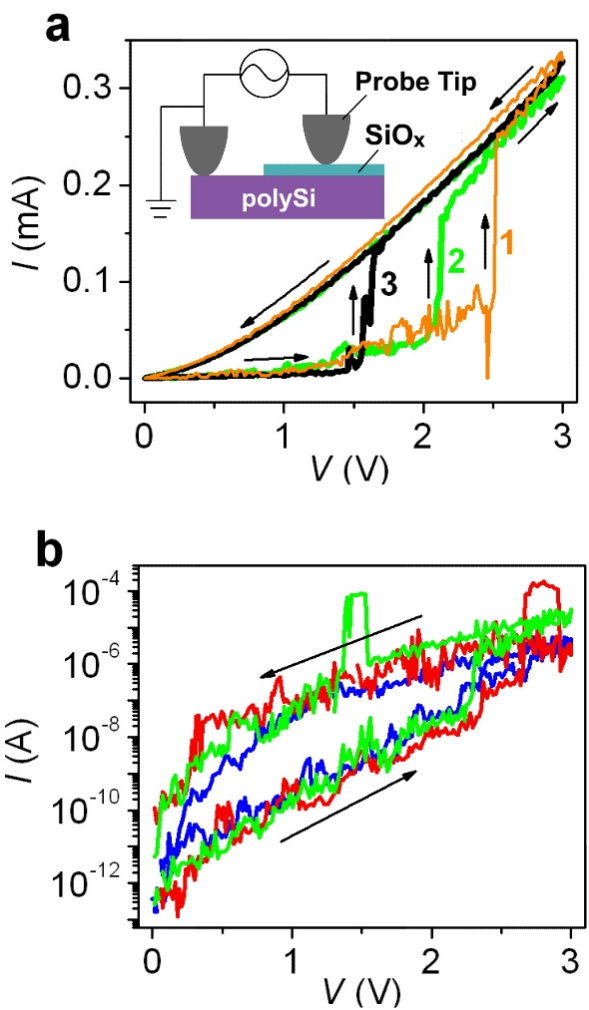

**Figure 8**. Electrical phenomena from surface native oxide. (a) Hysteretic *IV* curves from a surface native-oxide layer (1.5-3 nm thick). The numbers indicate the voltage-sweep order, and the arrows indicate the voltage-sweep direction. The inset is a schematic of the electrical setup. (b). Another series of hysteretic *IV* curves from a surface native-oxide layer, with the arrows indicting the voltage-sweep direction.

Finally, we tested a native-oxide layer (1.5-3 nm) atop a conducting polySi surface. Unlike previous systems where the $SiO_x$ layers are intentionally grown, here the $SiO_x$ layer is naturally formed on the Si surface in an ambient environment. This is relevant to some Si-electrode based molecular systems[46,47] in which, though $SiO_x$ may be not intentionally used, a native-oxide layer will inevitably be produced when the electrodes are fabricated.[40,44,48] We tested the electrical property of this native-oxide layer by directly landing two probe tips (tip diameters are ~ 20 um) on the polySi surface (see illustration in inset in Fig. 8a). We first formed a good ohmic contact between one tip and the polySi surface by a voltage sweep to a high value (e.g. > 5 V) to induce a hard electrical breakdown in the native-oxide layer. We then landed the other tip to a new location. Hence the electrical phenomena come from one tip-$SiO_x$-polySi interface. Fig. 8a shows a series of hysteretic *IV* loops from one of the interfaces we tested. Here the hysteresis differs from a set process in bistable NDR-like behavior (Fig. 6a) in that after each sweep loop, the subsequent sweep still starts with an OFF state, featuring the reproducibility and volatile property. The observed phenomena are very similar to those hysteretic behaviors observed in molecular systems at low voltage.[49,50] In addition to clear conductance-increase steps in Fig. 8a, Fig. 8b shows another series of current hysteretic *IV* curves with large current fluctuations from a native-oxide interface. Note that in both electrical phenomena, the native-oxide layer has not yet experienced a hard electrical breakdown, which is indicated from the reproducibility of the hysteresis and low conductance. Hence, the phenomena are likely to be charge-trap related. Hard electrical breakdown in the native-oxide layer is induced at a voltage above 3 V, after which the interface is permanently in an ohmic-contact state with no hysteretic behavior. For this reason, resistive switching and reproducible bistable NDR-like behavior as described before (Fig. 2) are not observed in native-oxide layers.

**Conclusion**

We have demonstrated various electrical phenomena including NDR-like behavior, resistive switching and current hysteresis intrinsic to a thin layer of $SiO_x$. These behaviors can largely mimic various electrical phenomena observed in molecules and other nanomaterials. The underlying cause for these effects is voltage-driven and high-electrical-field induced soft breakdown in the $SiO_x$ layer. In particular, this soft breakdown can be readily induced by unintended factors, such as defects in a $SiO_x$ surface, material-assisted local electrical-field enhancement and current local heating. Therefore, these results call for care when studying conduction in electronic systems with $SiO_x$ as a nominally passive component. The forming processes, behaviors, and mechanisms have been discussed in detail, providing a potential guide to distinguish electrical phenomena in molecules and nanomaterials of interest from those in $SiO_x$.

**Acknowledgement**

We thank Dr. J. Phillips, Rutgers University, for helpful discussions regarding stress distributions in $SiO_x$. D.N. acknowledges the support of the David and Lucille Packard Foundation. L.Z. acknowledges support from the Texas Instruments Leadership University Fund and National Science Foundation Award No. 0720825. J.M.T. acknowledges support from the Army Research Office through the SBIR program administrated by PrivaTran, LLC.

**Table of Contents Graphic**

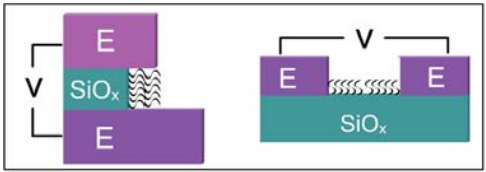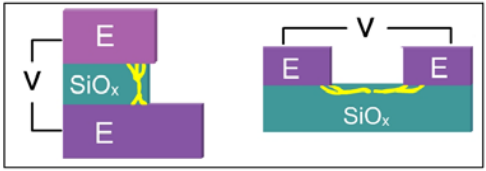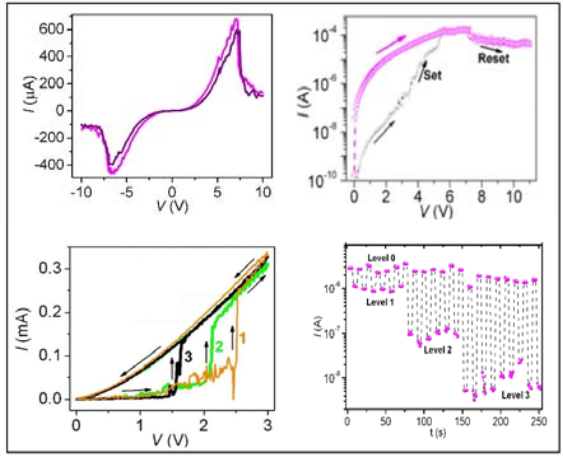